\documentclass[conference]{IEEEtran}
\ifCLASSINFOpdf
   \usepackage[pdftex]{graphicx}
\else
   \usepackage[dvips]{graphicx}
\fi
\usepackage[cmex10]{amsmath}
\begin{document}

\title{A novel method of generating tunable underlying network topologies for social simulation}

\author{
  \IEEEauthorblockN{
  Imre Varga\IEEEauthorrefmark{1}, 
  Andr\'as N\'emeth\IEEEauthorrefmark{1}, 
  Gergely Kocsis\IEEEauthorrefmark{1}
}\\

\IEEEauthorblockA{\IEEEauthorrefmark{1}Department of Informatics Systems and Networks, Faculty of Informatics, University of Debrecen, Hungary\\}
E-mail: {\tt\small \{varga.imre, kocsis.gergely\}@inf.unideb.hu}\\}

\maketitle

\begin{abstract}
We propose a method of generating different scale-free networks, which has several input parameters in order to adjust the structure, so that they can serve as a basis for computer simulation of real-world phenomena. The topological structure of these networks was studied to determine what kind of networks can be produced and how can we give the appropriate values of parameters to get a desired structure.
\end{abstract}

\IEEEpeerreviewmaketitle

amely az érzékelés, érzet, megismerés és megértés között zajló agyi folyamatok mérnöki informatikai modellezése

\section{Introduction}
Recently scale-free networks stand in the focus of research in several fields of science. Large scale of complex systems can be modeled using these structures (Social networks, ecological phenomena, metabolic processes, etc.) \cite{Newman}. In the last decade it also even turned out, that the human brain itself, as a network of correlated brain sites shares scale-free properties. Today this latter result is a pillar of modeling processes of the brain e.g. between cognition and understanding \cite{brain, cogi2}. As a basis of these researches scientists need methods to generate networks with properties similar to the object of their interest. In our case the main reason to create such networks is to investigate information spreading on them. In the literature a huge number of algorithms can be found such as preferential attachment \cite{Popularity, RMP74_47}, its extension with accelerated growth \cite{Accelerated}, fitness-driven process \cite{Fitness}, etc. However networks generated by these methods usually differ from the needed ones because the low amount of their parameters. There are efforts to create tunable methods, but they can change only few properties of the resulted networks \cite{Model, TunableCC, TunableCC2}. 
The goal of our research is to develop a network generation process, in which the input parameters can determine more structural properties. The far aim of this work is to create structures on which we can investigate spreading processes (using the model of Kocsis and Kun \cite{Infospread}). Namely we want to know how fast is the information spreading between agents of a specially created scale-free topology representing an online network. Since most of these networks share the same universal topological structure, it is likely that our method to generate networks like these, and the outcome of the research of the information spreading can be applied to heterogeneous human-computer networks as well, that seems to dominate future internet communication networks \cite{cogi1}. 
 

\section{The model}
History shows, that during the life-cycle of online social networks they go through more than one distinct phases. Namely after introduction, growing, and maturity sooner or later they start to decline. This phenomenon makes it reasonable to apply a two step generation process. 
Our {\it grow-and-destroy} network generation model has two stages. It is a mixture of the simple popularity-driven (BA) and the fitness-driven algorithm \cite{Model} extended by different attack methods. The generation of a growing network starts from a small connected network of $m_0$ initial nodes where each node has $2$ neighbors. The network is increased node-by-node. Each new node is linked to $m=m_0-2$ chosen from the existing ones. 
According to the BA algorithm the probability of connecting a new node $j$ to node $i$ is proportional to the actual number of connections $k_i$ of this node. Namely
\begin{equation}
p^{BA}_{ji}=\frac{k_i}{\sum\limits_{l=1}^{j-1}k_l}.
\end{equation}
In contrary in the fitness-driven model each node has a randomly assigned fitness value (a real value between $0$ and $1$) and the probability to join to node $i$ is given by the product of this fitness value and the number of existing links of the given node $k_i$. Thus
\begin{equation}
p^F_{ji}=f_jp^{BA}_{ji}=f_j\frac{k_i}{\sum\limits_{l=1}^{j-1}k_l}.
\end{equation}
Each node is linked to the network using BA algorithm with probability $p$ and using fitness-driven method with probability $1-p$ so the probability of a node $j$ to be linked to an already connected node $i$ is
\begin{equation}
p_{ji}=pp^{BA}_{ji}+(1-p)p^F_{ji}.
\end{equation}
The result is a connected growing network. 

However in many cases (e.g. online social networks) nodes are removed from the network as time passes. Thus when the size of our network $N_0$ reaches a desired value it goes through a so-called attack method. Namely $N_a$ nodes (and their links) will be removed from the network effecting fundamental structural properties of it. We take into account three completely different scenarios of removing. Thanks to the {\it rich gets richer} property \cite{Popularity} of preferential attachment, older nodes of the network (who were connected to the network earlier) can have more connections then younger ones. As a result of this the properties of a network are different if older nodes die out first or if recently attached, instable parts of the system are removed first or if the removing probability does not depends on number of connections.  

To catch this above described declining process of the network we used three different ways of attacking called central, peripheral and general attack. Central attack means that the removing probability of node $i$ is proportional to the value of its connections $k_i$. During peripheral attack nodes with a smaller number of neighbors have larger probability to remove while in the third case all nodes have the same chance to be removed. The strength of the attack $\eta$ can be defined by the fraction of the original and the removed number $N$ of nodes 
\begin{equation}
\eta=N_a/N_0=(N_0-N)/N_0.
\end{equation}

With the use of this complex grow-and-destroy network generation model one is able to produce a large variety of undirected scale-free network topologies.

\section{Results}
Based on the literature and our previous experiences we were interested in the following properties of the generated network topologies: 
\begin{itemize}
\item[{\it i})] Degree-distribution $P(k)$, i.e. what percentage of the nodes have a given number of neighbors. 
\item[{\it ii})] Average degree $\langle k\rangle$, namely the average number of links of each nodes. 
\item[{\it iii})] Cluster size distribution, where cluster means a small network which is not connected to others. 
\item[{\it iv})] Dominance of the giant component, i.e.  what percentage of nodes take place in the by far largest cluster of the network. 
\item[{\it v})] Average clustering coefficient $\langle C\rangle$ which describes how often neighbors of an average node are connected directly to each other as well.
\end{itemize}

Our model has five input parameters: system size ($N_0$), dominance of popularity-driven algorithm ($p$), number of links of a new node ($m$), type of attack process controlled by $p$ and the strength of attack ($\eta$). We wanted to know how the network properties depend on these parameters, explore this huge parameter space and determine the available regimes of network properties. In order to do this a large number of network generations have been carried out. 

\subsection{Degree-distribution}
\begin{figure}[!t]
\centering
\includegraphics[bb = 20 220 350 500, width=3.5in]{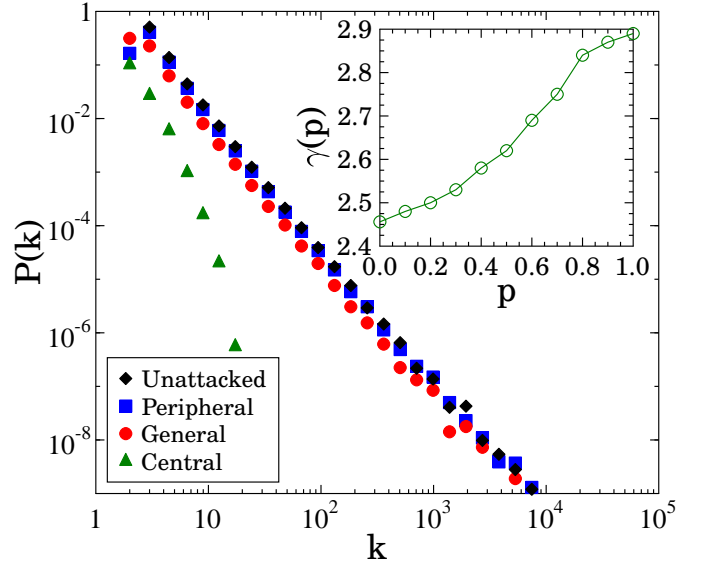}
\caption{Degree-distribution of original and differently attacked networks at attack rate $\eta=0.4$ ($N_0=10^6$, $m=3$). General and peripheral attack does not change degree distribution. Inset: The exponent of degree-distribution $\gamma$ as a monotonous function of $p$.}
\label{Pk}
\end{figure}

Most nodes of the generated networks have only a few neighbors. However there are some nodes with many connections to others. The degree-distribution $P(k)$ obeys power law behavior, since it is a scale-free network. The exponent $\gamma$ of the distribution is tunable, its value depends on the dominance of the BA-algorithm $p$ (Fig \ref{Pk} inset). The number of neighbors of a new node $m$ during the generation has no influence on the value of the exponent just shifts the curves. 
Data collapse can be achieved by rescaling by $2m^2$, thus the degree distribution can be written in this form
\begin{equation}
P(k) = 2m^2 k^{-\gamma(p)}.
\end{equation}
After the attack 
the degree-distribution can change. In the case of central attack first the exponent increases then the power law behavior disappears quickly, while in case of general and peripheral attack the power law dependence remains with almost the same exponent independently from the strength of attack $\eta$. (See Fig. \ref{Pk}.) 

\begin{figure}[!ht]
\centering
\includegraphics[bb = 30 220 350 500, width=3.5in]{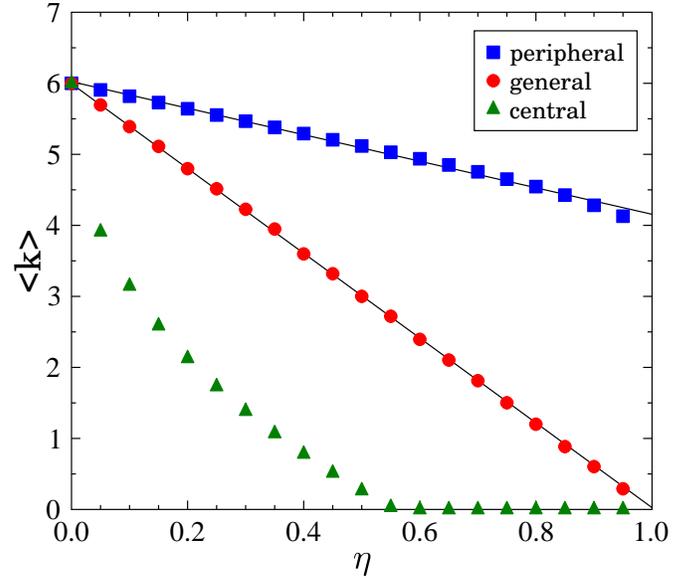}
\caption{The average degree of nodes $\langle k\rangle$ depends on the attack rate ($\eta$) at a certain value of $m$, however this dependence is determined by the type of the attack. ($N_0=10^6$, $m=3$, $p=0.1$)}
\label{avk}
\end{figure}
\subsection{Average degree}
The average degree (the average number of neighbors of a node) is determined only by the value of $m$ in the case of  unattacked networks. However when nodes are removed the average value of $k$ is changing. When we apply peripheral attack $\langle k\rangle$ is decreasing slowly, while in the case of central attack nodes loose their connections very fast as expected.
When general attack is used the reduction is between the former two cases. Except the cental case the value of the average number of connections $\langle k\rangle$ is decreasing almost linearly with the strength of attack this is presented on Fig. \ref{avk}. Not surprisingly the dominance of preferential attachment $p$ has no influence on the value of $\langle k\rangle$.

\subsection{Exponent of cluster size distribution}
\begin{figure}[!t]
\centering
\includegraphics[bb = 15 220 350 510, width=3.5in]{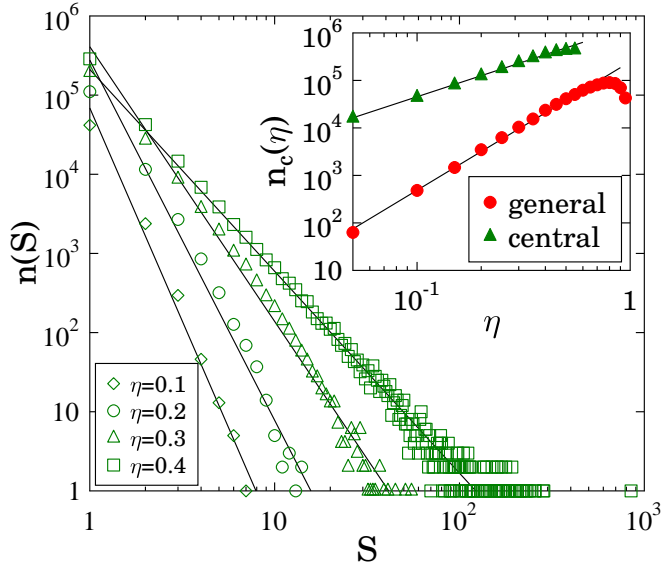}
\caption{The cluster size distribution at several attack rate from $0.1$ to $0.4$ in the case of central attack ($N_0=10^6$, $m=3$). The exponent $\tau$ is decreasing with $\eta$. Inset: Number of clusters in the system depends on attack rate $\eta$. The curves belong to central and general attack. Peripheral attack results in only a few clusters so the exponent can not be studied for that case.}
\label{ClusterSize}
\end{figure}

While the originally generated network is connected, as a reason of the attack process it breaks up to separate clusters. Usually there are some small separate groups of connected nodes and (in most of the cases) one so called giant component containing majority. Excluding this giant component the number of clusters $n(S)$ of size $S$ decreases as a power law (See Fig. \ref{ClusterSize}). By applying different attack methods it turned out that central attack results the lowest exponent and the most clusters. In contrary general attack leads to the lower number of clusters and to a higher exponent. Peripheral attack results more than one clusters only if we apply extreme attack rate. However such an extreme attack (above $0.9$) usually results in a so damaged network that makes it impossible to run simulations with practicable results. In all three cases the strength of the attack $\eta$ and the number of edges of new nodes $m$ have a large influence on the value of the exponent $\tau$. Consequently the size distribution of clusters (without the giant component) $n(S)$ can be cast to the form of
\begin{equation}
n(S)\sim S^{-\tau(\eta, m)}.
\end{equation}
The value of the exponent $\tau$ shows monotonous dependence of the attack rate $\eta$ (see Fig. \ref{ClusterSize}). The number of clusters in the system as a function of the attack rate $\eta$ has also power law functional form as it is presented on the inset of Fig. \ref{ClusterSize}. 

\subsection{Dominance of giant component}
\begin{figure}[!t]
\centering
\includegraphics[bb = 40 220 375 510, width=3.5in]{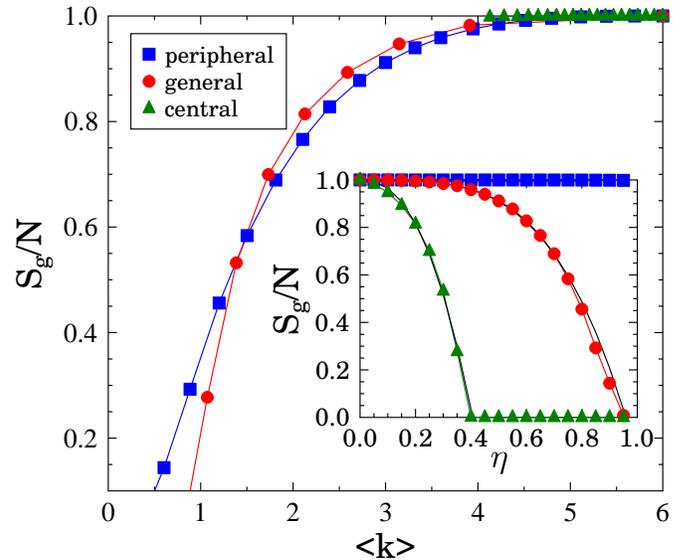} 
\caption{Average degree $\langle k\rangle$ of nodes has an influence on the size of the giant component, as the monotonous curves show. Namely more links result a larger giant component. Inset: The dominance of the giant component $S_g/N$ as a function of attack strength $\eta$ ($N_0=10^6$, $m=3$). Note that in a peripherally attacked network the giant component is always present {\it(blue)}, while if nodes with high value of degree are removed the giant component disappears if $\eta>0.4$ {\it(green)}, because the size of all clusters are in the same order of magnitude.}
\label{sgn}
\end{figure}

As it was presented above, as a result of the attack process clusters are present in the system, and most of the nodes form a giant cluster at low values of $\eta$. Increasing the attack strength this cluster looses its significance and become almost the same size as the other regular clusters. (See Fig. \ref{sgn} inset.) This process is highly similar to percolation, where the size of this giant component $S_g$ can be interpreted as a kind of order parameter of this second order phase transition \cite{Vicsek}. Using central attack the giant component disappears fast, while using peripheral attack the giant component stays always dominant almost independently of the strength of the attack. The transition point where this cluster is negligible can be shifted with the value of $m$. As expected, if more links are added at the generation to each nodes a larger giant component will be formed at a given $\eta$. However the dominance of the BA or the fitness driven algorithm described by the parameter $p$ does not effect the size.
The average degree $\langle k\rangle$ and the dominance of the giant component $S_g/N$ are not independent from each other. Larger $\langle k\rangle$ results a larger giant component as it is illustrated in the Fig. \ref{sgn}.

\subsection{Average clustering coefficient}
\begin{figure}[!t]
\centering
\includegraphics[bb = 0 220 350 500, width=3.5in]{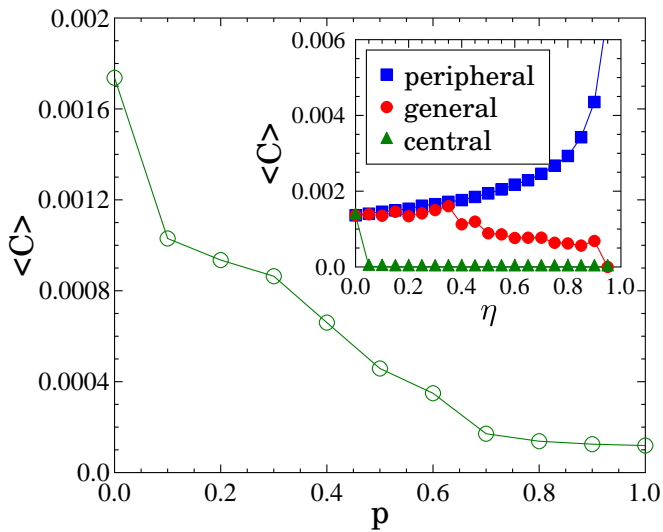} 
\caption{The average clustering coefficient as a function of dominance of BA algorithm, where $N_0=10^6$ and $m=3$. Inset: average clustering coefficient $\langle C\rangle$ strongly depends on the type of attack process and its strength ($p=0.1$).}
\label{ClustCoef}
\end{figure}

The value of the average clustering coefficient $\langle C\rangle$ is proportional to $m$ if $m>1$. The growing method also has an effect on $\langle C\rangle$. The neighbors of a randomly chosen node are linked together more often in a fitness-driven generated system then in a simple popularity-driven generated one (Fig. \ref{ClustCoef}). In both cases the increasing of the system size $N_0$ leads to a power law decay of the average clustering coefficient. The three attack methods affect the coefficient completely differently. General attack does not change the value of $\langle C\rangle$ for small values of $\eta$. In the case of central attack the average clustering coefficient decays very fast and becomes zero. Peripheral attack removes nodes with few connection thus the value of $\langle C\rangle$ is increasing with attack rate $\eta$. Inset of Fig. \ref{ClustCoef} illustrates these results.

\section{Conclusion}
In this work we developed a novel method of generating tunable underlying network topologies for social simulation.
As it was presented using our grow-and-destroy method one is able to generate scale-free network topologies with several tunable properties. Table \ref{limits} summarizes the minimum and maximum values of the studied quantities. These ranges of the output properties are not independent from each other, not all combinations of them can be generated. 
\begin{table}[ht]
\caption{Limits of the studied topological structure properties} 
\centering 
\begin{tabular}{ c c c c c } 
\hline\hline  \\ [0.5ex] 
\normalsize 2.4 &\normalsize  $<$ &\normalsize  $\gamma$           &\normalsize  $<$ &\normalsize  2.9   \\ 
\normalsize 0.1 &\normalsize  $<$ &\normalsize  $\langle k\rangle$ &\normalsize  $<$ &\normalsize  9.0   \\
\normalsize 2.5 &\normalsize  $<$ &\normalsize  $\tau$             &\normalsize  $<$ &\normalsize  7.0   \\
\normalsize 0.0 &\normalsize  $<$ &\normalsize  $\langle C\rangle$ &\normalsize  $<$ &\normalsize  0.1   \\
\normalsize 0.0 &\normalsize  $<$ &\normalsize  $S_g/N$            &\normalsize  $<$ &\normalsize  1.0   \\ [1ex] 
\hline 
\end{tabular}
\label{limits} 
\end{table}

Table \ref{influence} gives a short overview how the input parameters affect the output quantities. In the table $\uparrow$ indicates that with the increase of the input value the value of the output is also increasing and $\downarrow$ shows that an increasing input value results a decrease of the output. $\circ$ stands where the output does not depend on the input. $\updownarrow$ notes that both an increase or a  decrease is possible depending on the type of the attack.
\begin{table}[ht]
\caption{Influence of the input parameters to the output properties.} 
\centering 
\begin{tabular}{ c | c c c } 
\hline\hline   \\[-2ex] 
\normalsize                    &\normalsize  $p$          &\normalsize  $m$        &\normalsize  $\eta$        \\ 
\hline   \\[-1ex] 
\normalsize $\gamma$           &\normalsize  $\uparrow$   &\normalsize  $\circ$    &\normalsize  $\circ$       \\ 
\normalsize $\langle k\rangle$ &\normalsize  $\circ$      &\normalsize  $\uparrow$ &\normalsize  $\downarrow$  \\
\normalsize $\tau$             &\normalsize  $\circ$      &\normalsize  $\uparrow$ &\normalsize  $\downarrow$  \\
\normalsize $\langle C\rangle$ &\normalsize  $\downarrow$ &\normalsize  $\uparrow$ &\normalsize  $\updownarrow$\\
\normalsize $S_g/N$            &\normalsize  $\circ$      &\normalsize  $\uparrow$ &\normalsize  $\downarrow$  \\ [1ex] 
\hline 
\end{tabular}
\label{influence} 
\end{table}

Using our generation method we managed to generate different network families, e.g. one large connected network, a group of networks with or without a giant component, dense and connection-poor networks, etc. In our future research we plan to investigate information spreading on these generated network topologies. As an update of the generation method we would like to reproduce the so called ,,community of communities'' structure, or to take into account other properties during the generation phase such as the similarity of nodes \cite{PopSim}. Based on the results of investigation existing heterogeneous human-computer networks it would be also interesting to reproduce them using our method. 

\section*{Acknowledgment}The publication was supported by the T\'AMOP-4.2.2.C-11/1/KONV-2012-0001 project. The project has been supported by the European Union, co-financed by the European Social Fund.

\bibliographystyle{IEEEtran}
\bibliography{CogInfoCom2013}

\end{document}